\documentclass[runningheads]{cl2emult}

\usepackage{makeidx}  % allows index generation
\usepackage{graphicx} % standard LaTeX graphics tool
\usepackage{subeqnar} % subnumbers individual equations
\usepackage{multicol} % used for the two-column index
\usepackage{cropmark} % cropmarks for pages without
\usepackage{eso}      % placeholder for figures
\makeindex            % used for the subject index
\begin{document}
\title*{Cosmic shear surveys}
\toctitle{Cosmic shear surveys}
%\protect\newline in the Particle Deflection Plane}
% allows explicit linebreak for the table of content
%
%
\titlerunning{Cosmic shear surveys}
% allows abbreviation of title, if the full title is too long
% to fit in the running head
%
\author{Yannick Mellier\inst{1,2}
\and Ludovic van Waerbeke\inst{1}
\and Roberto Maoli\inst{3,1,2}
\and Peter Schneider\inst{4}
\and Buvnesh Jain\inst{5}
\and Francis Bernardeau\inst{6}
\and Thomas Erben\inst{1,2,7}
\and Bernard Fort\inst{1}
}
\authorrunning{Mellier et al.}
% if there are more than two authors,
% please abbreviate author list for running head
%
%
\institute{IAP, 98bis Boulevard Arago 75014 Paris, France,
\and Observatoire de Paris, DEMIRM, 61 avenue de l'Observatoire, 75014 Paris, France,
\and Dipartimento di Fisica, Universit\`a di Roma ``La Sapienza'', Italy 
\and Universitaet Bonn, Auf dem Huegel 71, 53121 Bonn, Germany
\and John Hopkins University, Dept. of Physics, Baltimore MD21218, USA
\and SPhT, CE Saclay, 91191 Gif-sur-Yvette Cedex, France
\and MPA, Karl-Schwarzscild Str. 1, 85748 Garching, Germany
     }

\maketitle              % typesets the title of the contribution

\begin{abstract}
Gravitational weak shear produced by large-scale 
structures of the universe induces a correlated ellipticity distribution 
 of distant galaxies.  The amplitude and evolution with angular scale 
of the signal depend on cosmological models and can be inverted in 
order to constrain the power spectrum and the cosmological parameters.
 We present our recent analysis of 50 uncorrelated VLT fields and the
  very first constrains on ($\Omega_m,\sigma_8$) and the nature 
  of primordial fluctuations based on the join analysis of present-day 
 cosmic shear surveys.
\end{abstract}

\section{Motivations}
The deformation of light beams produced by gravitational tidal fields 
  is responsible for the 
   cosmological weak lensing signal (or cosmic shear) produced by large-scale 
structures of the universe. The statistical properties of the 
  gravity-induced convergence, $\kappa$
  (the projected mass density) and  
 shear, $\gamma$ (the distortion) primarily depend on the 
  normalization of the  power spectrum of
  mass density fluctuations, $\sigma_8$, and on the density parameters, $\Omega_m$, and can be used to 
  constrain cosmological scenarios. 
 Bernardeau et al \cite{bernardeauetal} showed that the sensitivity of cosmological models 
  to these quantities is well described by the variance and the skewness of $\kappa$
   averaged over the angular scale 
 $\theta$, $\langle \kappa(\theta)^2 \rangle^{1/2}$ and 
  $s_3(\theta)$
\begin{equation}
\langle \kappa(\theta)^2 \rangle^{1/2} \approx \ 0.01 \ \sigma_8
\ \Omega_m^{\ 0.75} \ z_s
^{\ 0.8}
\left({\theta \over 1^o}\right)^{-(n+2)/3} \ ,
\end{equation}
\begin{equation}
s_3(\theta) \approx \ 40 \ \Omega_m^{\ -0.8} \ z_s^{\ -1.35} \ '
\end{equation}
 where $z_s$ is the redshift of sources. Hence, a cosmic shear survey which would focus on the
  measurement of the variance and the skewness of $\kappa$ should  recover both $\Omega_m$ and
  $\sigma_8$ independently.\\
Although the gravitational convergence is very weak, on angular scales smaller than 10 arc-minutes it 
  is enhanced by the the non-linear gravitational 
structures, which increase  the lensing signal by a significant amount
 (see Jain \& Seljak 1997, \cite{js}). 
  On those scales,  the cosmological weak lensing can already be measured from the gravity-induced 
 ellipticity of galaxies (the shear) even with ground-based telescopes. In fact, as
 it is shown by van Waerbeke et al 1999 (\cite{vwetal99}) and in Table \ref{sizesurvey}, one needs to cover about 
 one deg$^2$ up to $I \approx 24.5$ in order to measure cosmic shear on small scales. 
\\
Four teams recently presented first results. The most recent work was performed by Maoli et al 2000
(\cite{mvwmetal}) using VLT/FORS1 and has been used jointly with other surveys to explore 
  cosmological models. This work is summarized below.

\begin{table}[t]
\caption{\label{sizesurvey}Expected signal-to-noise ratio on the
the variance and the skewness of the convergence for two 
  cosmological models.  In the first
column, the size of the field of view (FOV) is given. The
signal-to-noise
 ratio is computed from the simulations done by
van Waerbeke et al  (1999). }
\begin{center}
\begin{tabular}{|l|c|c||c|c|}
\hline
\multicolumn{5}{|c|}{$z_s=1$, Top Hat
Filter , $n=30$ gal.arcmin$^{-2}$} \\ \hline
\hline
  FOV &
\multicolumn{2}{|c|}{ {S/N Variance}} &
\multicolumn{2}{|c|}{ {S/N Skewness}} \\
\cline {2-5}
 (deg.$\times$deg.) &  $\Omega_m=1$  &
 {$\Omega_m=0.3$}  &
 {$\Omega_m=1$}  &
 {$\Omega_m=0.3$}  \\ \hline
 1.25$\times$1.25  &  7  &  5 &  1.7 &   2 \\ \hline
 2.5$\times$2.5  & 11  & 10 & 2.9 &  4 \\ \hline
 5$\times$5  & 20  & 20 & 5 &  8\\ \hline
 10$\times$10  & 35 & 42 & 8 &  17\\ \hline
\end{tabular}
\end{center}
\end{table}

\section{Description of the VLT survey}
The VLT sample is defined in order to get a large number of fields 
 separated from each other by an angular distance as large as possible.
 This criterion enables us to minimize the cosmic variance.
 The selection of the field sample is optimized as follows:   
\begin{itemize}
\item no stars brighter than 8th magnitude inside a circle of 1 degree around
the FORS field, and no stars brighter than 14th magnitude inside the
FORS field, in order to avoid light scattering;
\item no extended  bright galaxies in the field. Their extended halo
 may contaminate the shape of galaxies located nearby;
\item neither rejection of over-dense regions, where clusters or groups of
galaxies could be present, nor primarily selection 
towards empty fields. Otherwise, the sample could be biased toward under-dense
regions with systematically low value of the convergence;
\item angular separation between each pointing larger than 5 degrees in
order to minimize the correlation between fields;
\item field must be galactic latitudes lower than $70^{\circ}$ in order to
get  enough stars per field  for the PSF correction.
\end{itemize}
We selected 50 FORS1 fields, each covering 6.8'$\times$6.8'.
The total field of view is 0.64 deg$^2$ and the pointings are randomly
 spread over more than 1000 deg$^2$. So far, this is the largest 
  sample of uncorrelated fields used for cosmic shear analysis.
\\
 The observations were obtained
with FORS1 on the VLT/UT1 (ANTU) at the Paranal Observatory in 
  I-band only.  They were carried out in service mode  which turns out to be 
 perfectly suited for our program. All the exposures have 
 a seeing between 0.55" and 0.80" with a median value at 0.64".
 The total exposure time per field is 36 minutes. 
  It has been computed in order to reach
$I=24.5$, which  corresponds to a galaxy number density per fields of about
30$\pm$ 10 gal.${\rm arcmin}^{-2}$. At this depth, 
  the expected average redshift of the
lensed sources is $\langle z\rangle \approx 1$. Note that thanks to 
the service mode observation, the VLT sample provides the most homogeneous 
  sample we have. From this sample, we extracted 76,000 galaxies. 
 Due to the severe selection criteria used for cosmic shear, the
final sample only has 50,000 galaxies.

\section{Results of cosmic shear experiments}
\begin{figure}
\centering
\includegraphics[width=0.7\textwidth]{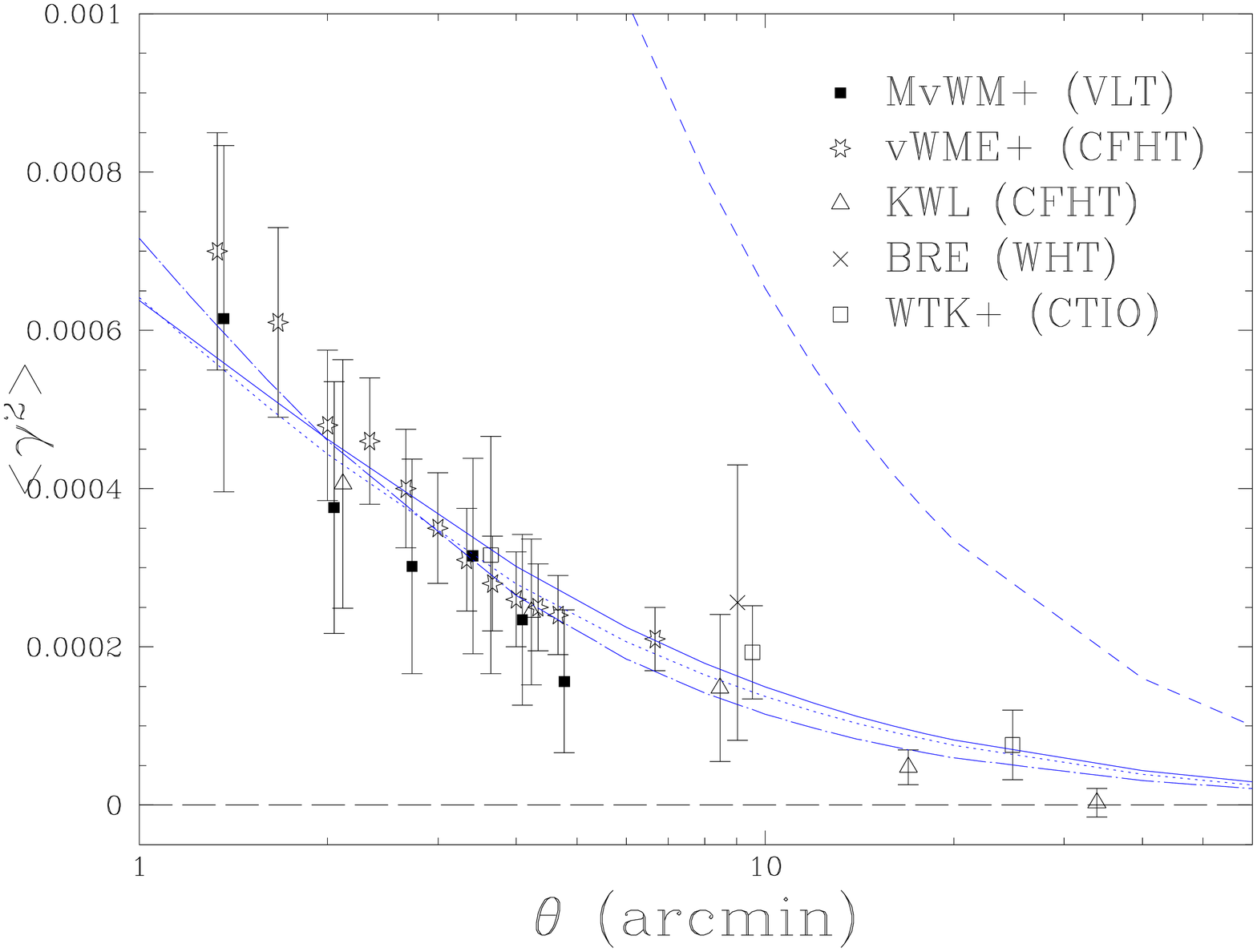}
\caption[]{
$\left<\gamma^2\right>$ as function of the
angular scale ($\theta$ is the angular diameter of a circular top-hat).
The filled squares are the VLT data.
 The other detection are \cite{vwmeetal} (vWME+),
\cite{baconetal1} (BRE), \cite{wittetal} (WTK) and \cite{mvwmetal} (KWL).
They are compared
to current cosmological models assuming a standard galaxies  redshift
distribution  with $z_0=0.8$. For all models
we choose a CDM power spectrum with $\Gamma=0.21$, and
$\Omega_0=1$, $\Lambda=0$, $\sigma_8=1$  (short dash);
$\Omega_0=0.3$, $\Lambda=0$, $\sigma_8=1.02$ (dot long-dash);
$\Omega_0=1.0$, $\Lambda=0$, $\sigma_8=0.6$ (dot) and
$\Omega_0=0.3$, $\Lambda=0.7$, $\sigma_8=1.02$ (solid). The models have
been computed using the non-linear evolution of the power spectrum
given by \cite{picdot96}.}
\label{shearfigmodel}
\end{figure}

So far, four teams have completed a cosmic shear survey.
 Each of them observed different fields
of view and used different
 instruments and techniques to get and to analyze the data
 (see Table \ref{summarysurvey}).
 The  CFHT and VLT surveys  reported in van Waerbeke et al
 \cite{vwmeetal} and  
 Maoli et al  \cite{mvwmetal} respectively 
 consist in two  independent
  data sets. We used them also to cross-check
our results and to explore the reliability of our
 corrections of  systematics.  
 The VLT sample complements our CFHT data which has the same depth,
  covers a much larger area ( 1.7 deg$^2$) but only contains 5 uncorrelated fields.  
 The use of both set of data simultaneously represents 75\% of the total number of fields 
 and 40\% of the total area covered by all cosmic shear surveys. 
\\
A description of the five surveys is summarized  in
Table \ref{summarysurvey} and the results are 
  in Fig. \ref{shearfigmodel}.
  The most striking
 feature on this plot is
  the remarkable similarity of the results in the range 1' to 10' .
This is a very strong point which validates the
  detection and guarantees that they are reliable and  robust, despite
   concerns about systematics.

\begin{table}
\caption{\label{summarysurvey}Summary of the 5 cosmic shear surveys
 completed so far. The CFHT
data were obtained with the UH8K and CFH12K CCD cameras. The R and I
limiting magnitudes enables us to estimate of the redshift of
the sources, which should be around one. }
\begin{center}
\begin{tabular}{|l|c|c|c|c|}
\hline
Reference & Telescope & Lim. Mag. & FOV & Nb. fields \\
 \hline
van Waerbeke et al \cite{vwmeetal} & CFHT & I=24 & 1.7 deg$^2$ & 5 \\
Wittman et al \cite{wittetal} & CTIO & R=26 & 1.5 deg$^2$ & 3 \\
Bacon et al \cite{baconetal1}& WHT & R=24 & 0.5 deg$^2$ & 13 \\
Kaiser et al \cite{kaiseretal}  & CFHT & I=24 & 1.0 deg$^2$ & 6 \\
Maoli et al  \cite{mvwmetal}& VLT-UT1 & I=24 & 0.5 deg$^2$ & 45 \\
\hline
\end{tabular}
\end{center}
\end{table}

\section{Cosmological interpretation of cosmic shear signal}
\begin{figure}
\centering
\includegraphics[width=0.7\textwidth]{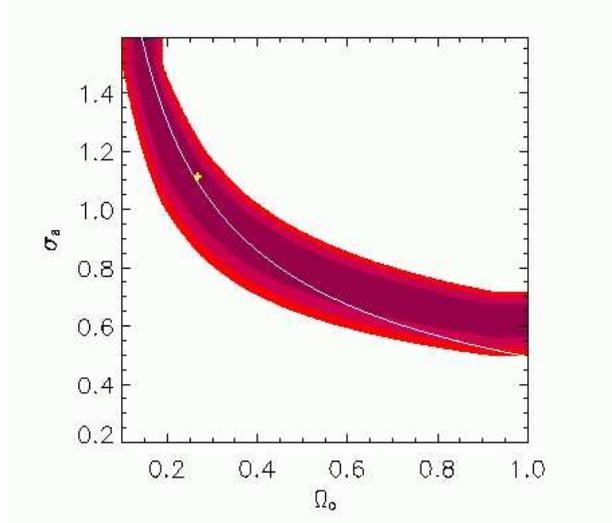}
\caption[]{The $\Omega_0$-$\sigma_8$ constraint derived
from combined cosmic shear surveys. The three grey areas define
the 1, 2 and
3-$\sigma$ limits. The cross indicates the position of the best fit
at $\Omega_0=0.26$ and $\sigma_8=1.1$. The solid line shows
the local cluster abundance best fit (\cite{pierpaoli}).
The latter and the cosmic shear constraints
have similar shape and match very well.
}
\label{chi2model}
\end{figure}

The results plotted in Figure \ref{shearfigmodel} confirm that the 
 Standard CDM predictions are incompatible with 
   most observations, including cosmic shear.  In contrast, cosmic shear 
   predictions 
  of most realistic  cluster normalized models  
  are all satisfactory, at least on scales
ranging from 0.5 to 10 arc-minutes. It is therefore 
   interesting to explore more thoroughly 
 a large set of models in a ($\Omega_0$,$\sigma_8$) space by using the 
  five cosmic shear results simultaneously.  The full sample 
  contains 75 uncorrelated fields and covers 5.5 deg.$^2$, so it  
 can already provide reliable informations.
\\
Since the five samples are independent, each provides one single 
 measurement point to perform a simple $\chi^2$ minimization in
the ($\Omega_0$,$\sigma_8$) plane. From each sample 
we choose only one point corresponding to the angular scale 
  where the signal has the best signal-to-noise, taking care to 
 discard the large scale measures, 
  as they are likely affected by finite size effects 
 (see Szapudi \& Colombi 1996,\cite{SC96}).
We extracted five triplets containing the
scale, the variance and the 1-$\sigma$ error,
($\theta_i$, $\gamma^2(\theta_i)$,$\delta\gamma^2(\theta_i)$) 
 reported on Figure \ref{shearfigmodel}) and
computed:

\begin{equation}
\chi^2={\displaystyle \sum_{i=1}^5 \left[{\gamma^2(\theta_i)-
\langle\gamma^2\rangle_{\theta_i} \over \delta\gamma^2(\theta_i)}\right]^2},
\end{equation}
where $\langle\gamma^2\rangle_{\theta_i}$ is the predicted variance for a given
cosmological model. We  computed it for 150 models inside
the box $0 <\Omega_0<1 $ and $0.2<\sigma_8<1.4$, with $\Gamma=0.21$,
$\Lambda=0$, and $z_0=0.8$.
 The result is given in Figure \ref{chi2model}.
The grey scales indicates the 1, 2 and 3-$\sigma$ confidence level
contours. We fitted the best models by the empirical
law:
\begin{equation}
\sigma_8\simeq 0.59 ^{+0.03}_{-0.03}\ \Omega_0^{-0.47}
\end{equation}
in the range $0.5<\theta<5$ arc-minutes which is found to be  is in good agreement
with \cite{js} who predicted $\sigma_8\propto \Omega_0^{-0.5}$ at
non-linear scales. Moreover, this is very close to the cluster normalization
constraints given in \cite{pierpaoli} (for closed models and $\Gamma=0.23$):
\begin{equation}
\sigma_8\simeq 0.495 ^{+0.034}_{-0.037}\ \Omega_0^{-0.60} \ 
\end{equation}
 although the two methods are totally independent.
 The interpretation of the remarkable agreement between the 
  cluster abundance and the cosmic shear analysis may be the following.
    The empirical law found from cluster abundance
  closely follows theoretical expectation of a Gaussian 
   initial density fluctuations field   
  (White et al 1993, \cite{WEF93}).  Since the amplitude of cosmic shear
 signal on scales smaller than 10 arc-minutes mainly probes non-linear mass
 density contrast like groups and clusters, the similarity between both
 empirical laws strengthens the assumptions that mass density fluctuations
 grew from a Gaussian field.
\\
Although encouraging, our interpretation of cosmic shear results depends
 on critical shortcomings. We only have five independent
data points spread over a rather small angular scale and we do not have 
  serious estimates on the redshift of the sources.
 Assuming they are at $z_0=0.8$ is a reasonable assumption
 (\cite{cohenetal}), but it is still
uncertain and needs further confirmations. 
  We also neglected the cosmic variance in the
error budget of the cosmic shear sample.  It does not
affect  the VLT data which contain 50 uncorrelated fields
and, likely, the Bacon et al (\cite{baconetal1}) observations 
  (because they estimated the cosmic
variance using a Gaussian field hypothesis). The
 three other measures are probably more affected, although numerical 
  simulations
 indicate that cosmic variance should only
increase our error bars by less than a factor of two (see \cite{vwmeetal}).

\section{Conclusions}
\begin{figure}
\centering
\includegraphics[width=0.7\textwidth]{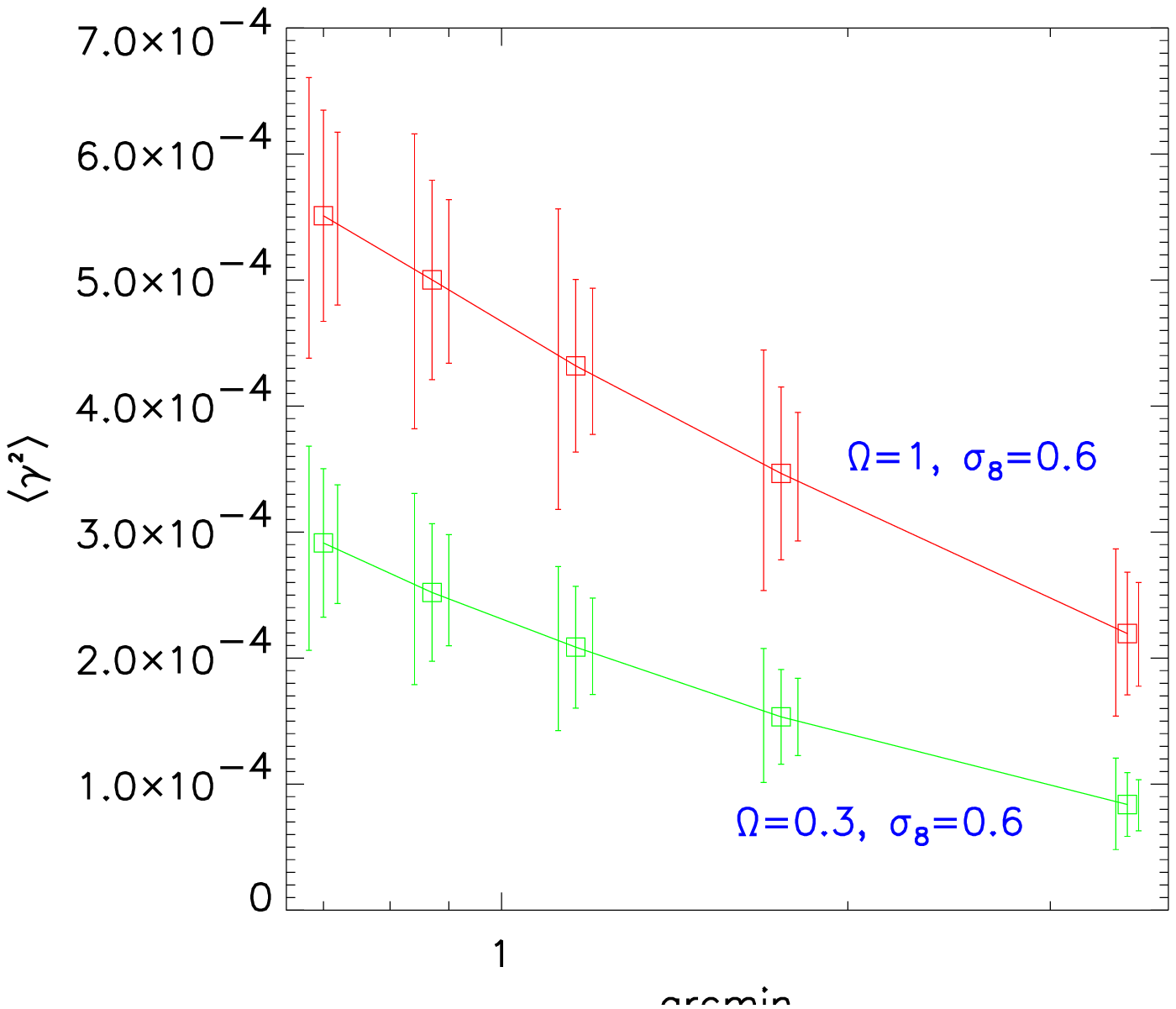}
\caption[]{Illustration of the signal-to-noise of the variance 
of the shear for 100, 150 and 300 VLT/FORS1 fields. One can see
  that for 300 hundred fields, the two models will be 
disentangled with a 3$\sigma$ confidence level.
}
\label{sumnbvlt}
\end{figure}

Although cosmic shear surveys started less than two years ago, they  
 went incredibly fast to provide consistent measurements on small 
scales. The study discussed in this proceeding goes even further.
 It both shows the important immediate potential of cosmic shear for cosmology 
and the fact that FORS1 in service mode is one of the best 
  instrument for this project. It enables to get a homogeneous 
 data set on a very large sample of uncorrelated fields.
\\
On Figure \ref{sumnbvlt}, we have simulated the amount of data one would need in order to
increase the signal-to-noise ratio by a factor of 3.  It turns out
that with 300 FORS1 fields obtained in service mode (that is, 250 more
fields than what we got, or 160 hours of ANTU/FORS1 in service mode) 
 the separation between most popular model would be 
  striking.  If the VMOS instrument 
 (Le F\`evre et al 2000, \cite{lefevretal2000}) provides similar image quality, 
  one can imagine even more impressive results up to angular scales of 
  15 arc-minutes.  The join use of both CFHT and VLT data would therefore 
  be spectacular.  In particular,
the skewness of the convergence, which is insensitive to $\sigma_8$,
will appear as a very narrow vertical constraint
on Figure \ref{chi2model} therefore breaking
the $\Omega_0$-$\sigma_8$ degeneracy.

\section*{acknowledgements}
We thank the ESO staff in Paranal observatory for
the observations they did for us in Service Mode.
This work was supported by the TMR Network ``Gravitational Lensing: New
Constraints on
Cosmology and the Distribution of Dark Matter'' of the EC under contract
No. ERBFMRX-CT97-0172. 
 We thank the TERAPIX data center for providing its facilities
for the data reduction of the VLT/FORS data. 
%\end{acknowledgements}

%INDEX%%%%%%%%%%%%%%%%%%%%%%%%%%%%%%%%%%%%%%%%%%%%%%%%%%%%%%%%%%%%%%%
\clearpage
\addcontentsline{toc}{section}{Index}
\flushbottom
\printindex
%%%%%%%%%%%%%%%%%%%%%%%%%%%%%%%%%%%%%%%%%%%%%%%%%%%%%%%%%%%%%%%%%%%%%

\end{document}